\documentclass[%
reprint,
superscriptaddress,
showkeys,
amsmath,amssymb,
pra,
floatfix,
]{revtex4-2}

\usepackage{defs}

\begin{document}

\title{Control of Spin Polarization through Recollisions}

\author{Stefanos~Carlström\,\orcidlink{0000-0002-1230-4496}}%
\email{stefanos@mbi-berlin.de}
\email{stefanos.carlstrom@matfys.lth.se}
\affiliation{Max-Born-Institut, Max-Born-Straße 2A, 12489 Berlin, Germany}
\affiliation{Department of Physics, Lund University, Box 118, SE-221 00 Lund, Sweden}

\author{Jan~Marcus~Dahlström\,\orcidlink{0000-0002-5274-1009}}%
\affiliation{Department of Physics, Lund University, Box 118, SE-221 00 Lund, Sweden}

\author{Misha~Yu~Ivanov\,\orcidlink{0000-0002-8817-2469}}%
\affiliation{Max-Born-Institut, Max-Born-Straße 2A, 12489 Berlin, Germany}
\affiliation{Department of Physics, Imperial College London, South Kensington Campus, SW72AZ London, United Kingdom}
\affiliation{Institut für Physik, Humboldt-Universität zu Berlin, Newtonstraße 15, 12487 Berlin, Germany}

\author{Olga~Smirnova\,\orcidlink{0000-0002-7746-5733}}%
\affiliation{Max-Born-Institut, Max-Born-Straße 2A, 12489 Berlin, Germany}
\affiliation{Technische Universität Berlin, Ernst-Ruska-Gebäude, Hardenbergstraße 36A, 10623 Berlin, Germany}

\author{Serguei~Patchkovskii}%
\affiliation{Max-Born-Institut, Max-Born-Straße 2A, 12489 Berlin, Germany}

\date{\today}

\begin{abstract}
  Using only linearly polarized light, we study the possibility of
  generating spin-polarized photoelectrons from xenon atoms. No net
  spin polarization is possible, since the xenon ground state is
  spin-less, but when the photoelectron are measured in coincidence
  with the residual ion, spin polarization emerges. Furthermore, we
  show that ultrafast dynamics of the recolliding photoelectrons
  contribute to an apparent flipping of the spin of the photoelectron,
  a process that has been completely neglected so far in all analyses
  of recollision-based processes. We link this phenomenon to the
  \enquote{spin--orbit clock} of the remaining ion. These effects arise
  already in dipole approximation.
  \begin{description}
  \item[Published in] \emph{Physical Review A} \textbf{108}(4), 043104 (2023),
    DOI: \href{https://doi.org/10.1103/PhysRevA.108.043104}{10.1103/PhysRevA.108.043104}
  \end{description}
\end{abstract}


\maketitle

\section{Introduction}

Generation of spin-polarized photoelectrons using intense circularly
polarized light has recently become a topic of great interest
\cite{Barth2014,Hartung2016,Trabert2018,Nie2021,Mayer2022}. Since the
rare gases commonly employed in strong-field ionization experiments
are spin-less in the ground state, linearly polarized light cannot
generate net spin polarization. In this article we show that when the
photoelectron is measured in coincidence with the final ion state, the
spin polarization approaches \SI{100}{\percent} in the individual
ionization channels (resolved on \(J\) and \(M_J\)). Furthermore, we
link the resulting spin polarization to the rescattering electron
imaging the ultrafast hole motion, providing an intuitive picture of
electron trajectories that contribute to an apparent spin flip of the
detected electron --- a signature of recollision-driven coupling between
continua with different spins. We find that the spin-flip recollisions
are very significant, and that we may exercise precise control over
the outcome. This effect, which has so far been overlooked, is
important in all recollision-based imaging techniques such as
laser-induced electron diffraction \cite{Zuo1996}, electron holography
\cite{Huismans2010}, and orbital tomography
\cite{Patchkovskii2006,Patchkovskii2007}.

This article is arranged as follows: section~\ref{sec:theory}
introduces the computational tools we employ in the calculations,
presented in
section~\ref{sec:calculations}. Section~\ref{sec:conclusions}
concludes the article.

\section{Theory}
\label{sec:theory}

Our method consistently treats multi-electron spin dynamics in strong
laser fields, and is thus suitable for our chosen target, xenon. It is
based upon the time-dependent configuration-interaction singles
(TD-CIS) \cite{Rohringer2006,Greenman2010PRA,Carlstroem2022tdcisI,
  *Carlstroem2022tdcisII}. The equations of motion (EOMs) describe the
time evolution of the amplitude \(c_0\) for the Hartree–Fock (HF)
reference state, and the particle orbital \(\contket*{k}\) emanating
from the initially occupied (time-independent) orbital \(\ket{k}\).
Below, we employ Hartree atomic units. Quantities appearing on one
side only are summed/integrated over. The different particle--hole
channels can couple via both the laser interaction and the Coulomb
interaction:
\begin{equation}
  \label{eqn:td-cis-eoms-simplified}
  \begin{aligned}
    \imdt c_0
    &=
      {\matrixel*{k}{\laserinteraction}{\cont{k}}}, \\
    \imdt\contket*{k}
    &=
      (-\orbitalenergy{k}+{\fock})\contket*{k} +
      {c_0\laserinteraction\ket{k}} -
      {\matrixel{l}{\laserinteraction}{k}}\contket*{l} \\
    &
      \hphantom{=}
      -{(\direct[lk]-\exchange[lk])\contket*{l}} -
      \lagrange{\cont{k}m}\ket{m},
  \end{aligned}
\end{equation}
where \(\orbitalenergy{k}\) is the eigenvalue of the initially
occupied orbital \(\ket{k}\). The Fock operator is defined as
\(\fock\defd \hamiltonian + \direct[mm] - \exchange[mm]\), with the
one-body Hamiltonian containing the interaction with the external
laser field,
\(\hamiltonian \defd p^2/2 + \nuclearpotential(\vec{r}) +
\laserinteraction\),
\(\laserinteraction\defd \fieldamplitude{t}\cdot\vec{r}\), and
\(\direct[cd]\) and \(\exchange[cd]\) are the \emph{direct} and
\emph{exchange interaction} potentials, respectively
\citeApp{app:coulomb-repulsion}. The Lagrange multipliers
\(\lagrange{\cont{k}m}\) in
Equation~\eqref{eqn:td-cis-eoms-simplified} ensure that
\(\contket*{k}\) at all times remains orthogonal to all initially
occupied orbitals \(\ket{m}\).

\begin{figure*}[htb]
  \centering
  \includegraphics{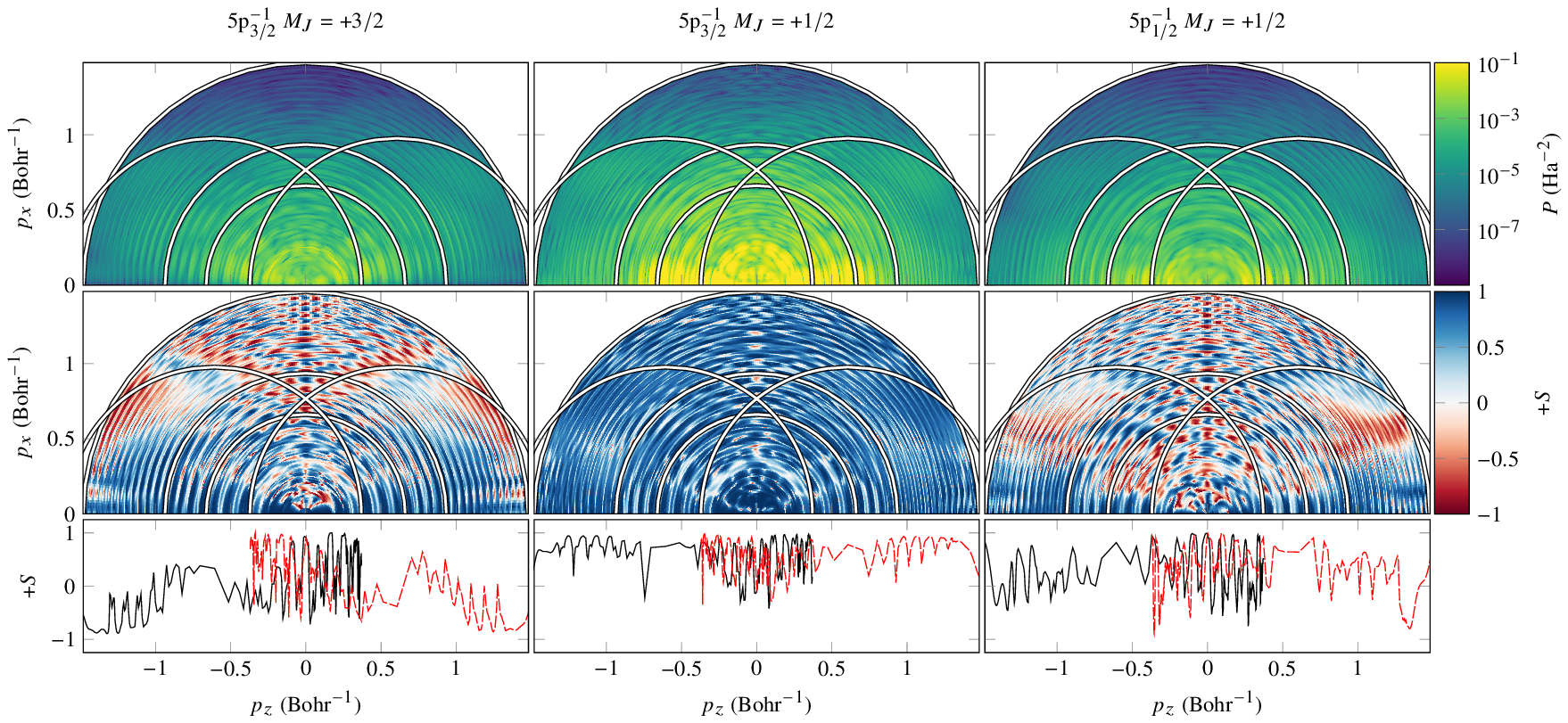}
  \caption{\label{fig:spin-polarization-angle-resolved} \emph{Top
      row}: Angularly resolved ATI spectra from xenon, correlated to
    different ionic channels. \emph{Middle row}: Spin polarization of
    the photoelectrons; for \(M_J<0\), the spin polarization is
    exactly the opposite of the positive case. Blue colour means
    excess of spin-up (spin-down) for \(+S\) (\(-S\)), and vice versa
    for red. The overlaid concentric circles indicate momenta at the
    detector corresponding to kinetic energies \(2\ponderomotive\),
    \(4\ponderomotive\), \(10\ponderomotive\), respectively. The
    offset circles mark the contribution of recolliding electrons with
    the maximum possible return energy \(\kinengmax\). \emph{Bottom
      row}: Lineouts along the \(\kinengmax\) circles in the middle
    row; the black solid (red dashed) line corresponds to initial
    ionization towards positive (negative) \(p_z\).}
\end{figure*}
To implement spin--orbit coupling, instead of resorting to the full
four-component Dirac--Fock treatment (RTDCIS \cite{Zapata2022}), we
rely on the phenomenological two-component treatment of
\textcite{Peterson2003}. It includes corrections due to
scalar-relativistic effects, and at the same time reduces the number
of electrons we need to treat explicitly. It replaces the scalar
potential \(\nuclearpotential\) by the relativistic effective core
potential (RECP), which models the atomic nucleus and the 1s--3d
electrons of xenon according to
\begin{equation}
  \label{eqn:recp}
  \pseudopotential(\vec{r}) =
  \scalarpotential(r) +
  \mu\sopotential_{\ell j}(r).
\end{equation}
The RECP allows us to identify effects associated with spin--orbit
dynamics by scaling the spin--orbit splitting as
\begin{equation}
  \label{eqn:so-scaling}
  \DEso(\mu) \approx \mu\DEso(1),
\end{equation}
where \(\DEso(1)\approx\SI{1.4}{\electronvolt}\) is the nominal spin--orbit
splitting of the ionic ground state at the CIS level; the dependence
is essentially linear in \(\mu\) \citeApp{app:scaling-so}.

The spin polarization is given by
\begin{equation*}
  S_I(E,\theta) \defd
  \frac{P_{I\alpha}(E,\theta)-P_{I\beta}(E,\theta)}{P_{I\alpha}(E,\theta)+P_{I\beta}(E,\theta)},
\end{equation*}
where \(P_{I\spin}(E,\theta)\) is the ion-, kinetic energy-, angle-, and
spin-resolved photoelectron distribution
\citeApp{app:photoelectron-spectra}.

\section{Calculations}
\label{sec:calculations}

We study above-threshold ionization (ATI) from xenon, with the
following ionization channels included: \channela{},
\(M_J=\pm3/2,\pm1/2\), and \channelb{}, \(M_J=\pm1/2\) \footnote{The quantum
  numbers \(J\) and \(M_J\) pertain to the states of the ion, whereas
  \(j\) and \(m_j\) label the initially occupied orbitals; for CIS
  from closed valence shells, \(J=j\) and \(M_J=-m_j\). In \(LS\)
  coupling, the \conf{p} spin-orbitals are labelled
  \(\conf{p}_{m_\ell}\spin\), where the possible values for \(m_\ell\) are
  \(0\), \(\pm\), and \(\spin=\alpha,\beta\) corresponding to
  spin-up/-down.}. Ionization from \conf{5s} and lower-lying orbitals
is strongly suppressed in the laser fields we consider
[\(\hbar\omega=\DEso(1)\) and
\(I_0=\SI{44}{\tera\watt\per\centi\meter\squared}\)]. The
\emph{spin-mixed} channels \channela{}, \(M_J=\pm1/2\) (formed from
linear combinations of \(\conf{p}_0\alpha\), \(\conf{p}_+\beta\) and
\(\conf{p}_0\beta\), \(\conf{p}_-\alpha\), respectively) are preferentially
ionized, since ionization in linearly polarized fields is dominated by
\(\conf{p}_0\) \cite{Perelomov1966SPJ}.

The weaker channels \channela{}, \(M_J=\pm3/2\) are expected to be
\emph{spin-pure}, since to form the orbitals \(\conf{5p}_{3/2}\),
\(m_j=\pm3/2\), the orbital- and spin-angular momenta must be maximally
aligned (\(\conf{p}_+\alpha\) and \(\conf{p}_-\beta\), respectively). Linearly
polarized electric fields preserve spin, and thus we expect that the
outgoing electron is spin-pure as well. However, the results of our
numerical simulations are surprising: only direct, on-axis
photoelectrons maintain their expected spin (see
Figure~\ref{fig:spin-polarization-angle-resolved}).  In contrast
electrons that have undergone recollision with the parent ion, and are
able to travel off-axis, exhibit substantial amounts of the opposite
spin.
\begin{figure}[htb]
  \centering
  \includegraphics{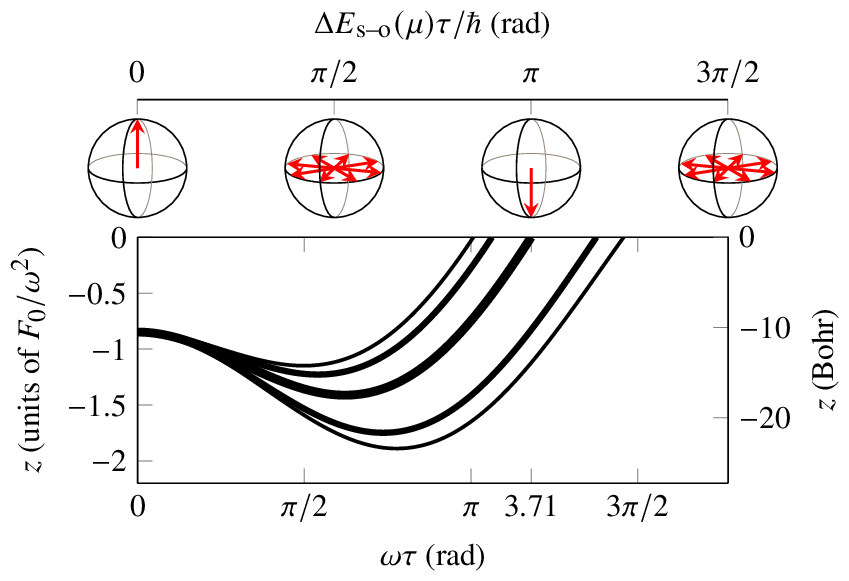}
  \caption{Sketch of the proposed spin-flipping mechanism; the
    electron moves along a rescattering trajectory with an excursion
    phase of \(\omega\tau\), which in the example is chosen such that the
    trajectory with maximum return energy \(\kinengmax\), shown as the
    heaviest line, matches half a revolution of the spin--orbit clock,
    illustrated as Bloch spheres. The classical trajectories start at
    the tunnel exit of a Coulombic barrier.}
  \label{fig:sketch}
\end{figure}

To explain this behaviour, we posit that this effect results from the
recollision of the returning electron off of the ion, which in the
spin-mixed channels has time-evolving spin
\cite{Barth2014,Mayer2022}. Directly after ionization, the ion has a
spin opposite that of the photoelectron, yielding vanishing spin
overall. If upon return, the electron finds an ion with a spin
different from that at time of ionization, inelastic scattering into
the spin-pure channels \channela{}, \(M_J=\pm3/2\) may contribute to
photoelectrons of opposite spin in these channels. Furthermore, this
apparent spin flip will predominantly occur when
\(\DEso\tau\approx\pi\), where \(\tau\) is the excursion time of the electron, see
Figure~\ref{fig:sketch} and the SI
\citeApp{app:classical-trajectories}. This dynamic corresponds to the
\emph{spin--orbit clock} in the ion undergoing half a revolution.

To investigate this hypothesis in a minimally invasive manner, we
tune the spin--orbit splitting \(\DEso\) by changing the value of \(\mu\)
in \eqref{eqn:recp}, while keeping all remaining parameters constant. We
then find
\begin{equation}
  \pi\approx\DEso(\mu)\tau = \frac{\DEso(\mu)}{\omega} \omega\tau =
  \frac{\DEso(\mu)}{\DEso(1)} \omega\tau,
\end{equation}
since we chose the photon energy to be in resonance with the nominal
spin--orbit splitting, \(\omega=\DEso(1)\). Using
\eqref{eqn:so-scaling}, we get
\begin{equation}
  \mu \approx \frac{\DEso(\mu)}{\DEso(1)} \approx \frac{\pi}{\omega\tau}.
\end{equation}
For electrons returning with maximal kinetic energy, \(\kinengmax\),
which return at \(\omega\tau\approx3.71\), we obtain
\(\mu\approx0.85\). It is easy to find those final momenta (combinations of
\(p_z\) and \(p_x\)) which result from trajectories recolliding with
\(\kinengmax\) \cite{Spanner2004}
\citeApp{app:classical-trajectories}; these are marked in
Figure~\ref{fig:spin-polarization-angle-resolved} with circles in the
forward (\(p_z>0\)) and backward (\(p_z<0\)) directions. If we take
lineouts of the spin polarization along these circles, we
predominantly measure the contribution of trajectories returning with
\(\kinengmax\) kinetic energy. The red streaks in
Figure~\ref{fig:spin-polarization-angle-resolved} that indicate the
opposite spin do not fall perfectly on the \(\kinengmax\) circle; this
is mostly due to the circle being derived for classical trajectories
with no potential present. The slight shift in momentum for the
apparent spin flips is a result of Coulomb focusing.

\begin{figure}[htb]
  \centering
  \includegraphics{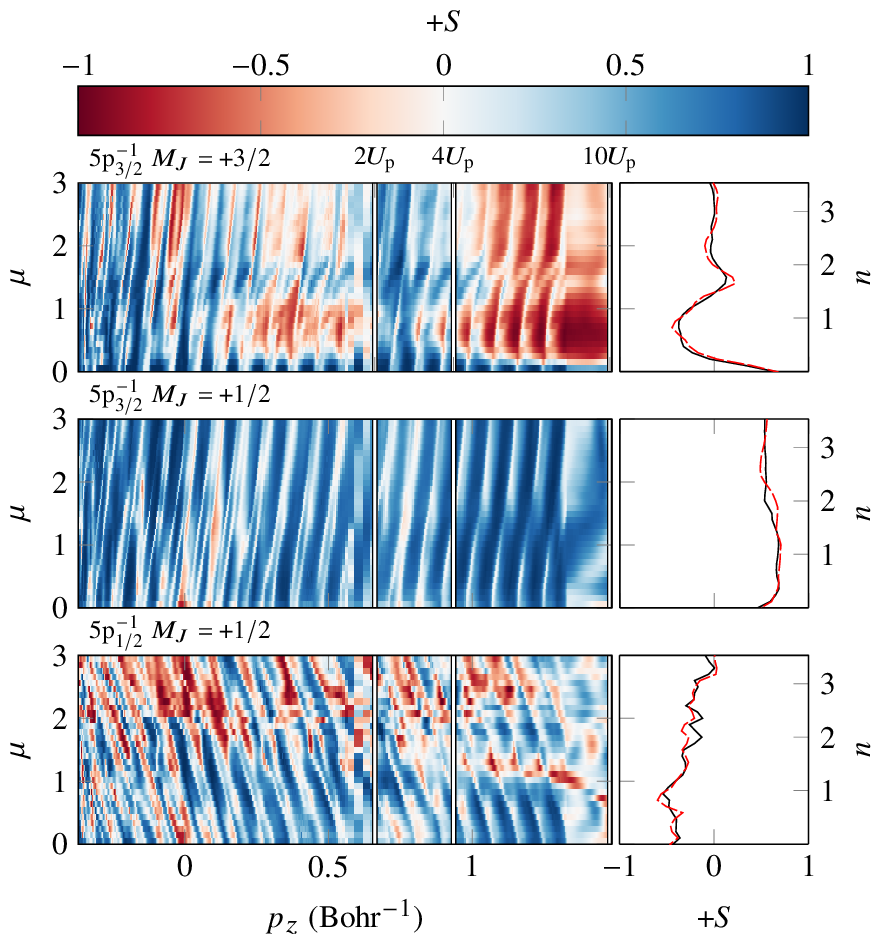}
  \caption{\label{fig:spin-polarization-3Up-lineouts} Lineouts along
    the \(\kinengmax\) circles in
    Figure~\ref{fig:spin-polarization-angle-resolved}, for a range of
    values of the spin--orbit scaling parameter \(\mu\) [see
    Equation~\eqref{eqn:so-scaling}], plotted as a function of
    \(p_z\); note that \(2\kineng = p_z^2+p_x^2\), and the lineouts
    are \emph{not} taken along constant \(p_x\). As in
    Figure~\ref{fig:spin-polarization-angle-resolved}, for \(M_J<0\),
    the spin polarization is exactly \(-S\). Each row corresponds to a
    specific ion channel; \emph{top row} the spin-pure channel
    \channela{}, \(M_J=+3/2\), \emph{middle row} \channela{},
    \(M_J=+1/2\), \emph{bottom row} \channelb{}, \(M_J=+1/2\). The
    \emph{left column} shows emission in the forward direction, i.e.\
    positive final \(p_z\) [due to the long pulse duration,
    \(\tau=\SI{15}{\femto\second}\), the spin polarization is almost
    symmetric about \(p_z=0\) \citeApp{app:lineouts-backwards}]. The
    \emph{right column} shows the integrated spin polarization along
    the lineout, from \(p^2/2 = 4\ponderomotive\) to \(\pmax\); the
    solid, black line corresponds to the forward emission, and the
    dashed red to the backward emission. The right ordinate indicate
    the expected positions corresponding to
    \(n\approx\mu\omega\tau/\pi\) half-revolutions of the spin--orbit clock, for
    \(\omega\tau\approx3.71\) (see main text).}
\end{figure}
We thus expect large amounts of opposite spin in the high-rescattering
region (\(\abs{\kineng}>4\ponderomotive\)), for \(\mu\approx0.85\), since the
\emph{spin--orbit clock} has undergone half a revolution, by the time
electron returns. In Figure~\ref{fig:spin-polarization-3Up-lineouts}
we see that this is indeed the case, in the spin-pure channels
\channela{}, \(M_J=\pm3/2\). Generalizing this argument, for
\(\mu\approx\frac{n\pi}{\omega\tau}=0.85, 1.69, 2.54, 3.40, 4.23, 5.07, ...\) we expect
to see enhancement and suppression of the opposite spin for odd \(n\)
and even \(n\), respectively.

It is also interesting to note that for \(\mu=0\), the photoelectrons in
the spin-pure channel are spin-pure as well. In this case, the period
of the \emph{spin--orbit clock} is
\(\frac{2\pi}{\DEso(0^+)}=+\infty\), and the hole remains forever in its
initial spin state, preventing any opposite spin appearing in the
spin-pure channels.

\begin{figure*}[htb]
  \centering
  \includegraphics{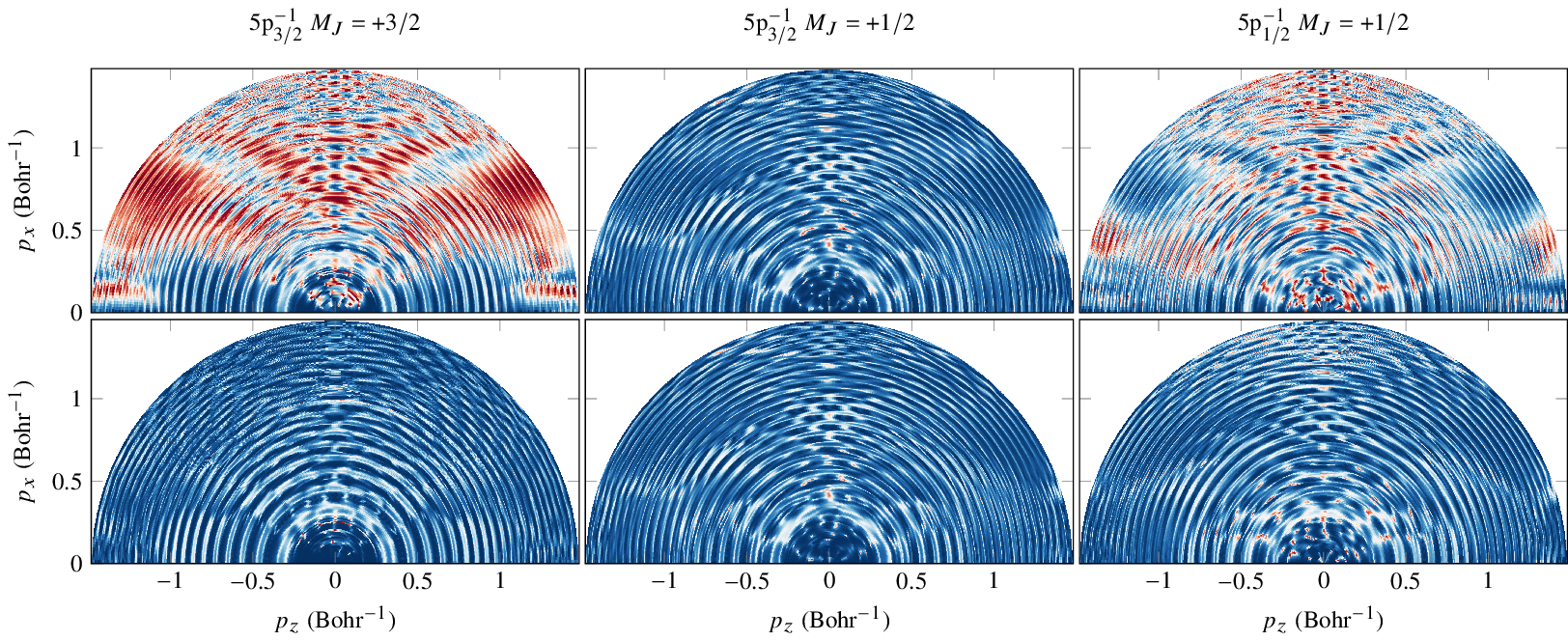}
  \caption{\label{fig:spin-polarization-kill-coulomb} Effect on the
    spin polarization when gradually removing the Coulomb
    electron--electron repulsion interaction from the
    EOMs~\eqref{eqn:td-cis-eoms-simplified}. \emph{Top row}:
    \(\exchange[lk]\) dropped; \emph{bottom row};
    \(\exchange[lk],\direct[lk]\) dropped. See the second row of
    Figure~\ref{fig:spin-polarization-angle-resolved} for the case
    where all terms in the Hamiltonian are included.}
\end{figure*}
To further explore the proposed mechanism, we selectively remove the
Coulomb repulsion interaction from the EOMs
\eqref{eqn:td-cis-eoms-simplified}; first we exclude exchange-type
interactions between ionization channels by dropping the
\(\exchange[lk]\) term, and then the direct-type interchannel
interactions \(\direct[lk]\). Dropping the self-interaction correction
\(\exchange[mm]\) does not influence the spin polarization appreciably
\citeApp{app:remove-Kmm}. The intrachannel interactions
\(\direct[mm]\) must remain, since otherwise the problem would reduce
to a hydrogenic one with a bare xenon nucleus. We compare these
instrumented calculations with the full Hamiltonian in
Figure~\ref{fig:spin-polarization-kill-coulomb}. As we see in the
figure, the largest effect is the removal of \(\direct[lk]\), which is
the only term of the three which is long-range (\(\exchange\), being
traceless, decays at least as quickly as \(r^{-2}\)). We also note
that the removal of \(\exchange[lk]\) quantitatively changes the
angular distribution of the spin polarization, even enhancing it,
which suggests that \(\exchange[lk]\) actually works \emph{counter} to
the proposed mechanism.

We now understand the mechanism leading to the opposite spin in the
spin-pure channel better: The hole in this channel is also spin-pure,
and as such does not undergo any spin oscillation in the
\emph{spin--orbit clock}. However, the holes in the other channels are
spin-mixed, and the \emph{spin--orbit clock} oscillates with the period
\(\Tso\defd\frac{2\pi}{\DEso(\mu)}\). After rescattering at the right
moment, we may observe opposite spins due to inelastic scattering into
\channela{}, \(M_J=\pm3/2\). Removing \(\direct[lk]\) from the EOMs
\eqref{eqn:td-cis-eoms-simplified} suppresses inelastic scattering,
and thus precludes any transfer of spin between channels, as we see in
Figure~\ref{fig:spin-polarization-kill-coulomb}. This mechanism can be
semi-quantitatively investigated by considering the explicit time--spin
dependence of \(\direct[lk]\) and \(\exchange[lk]\) in \(LS\)
coupling, where the orbitals \(l\) and \(k\) change their spin with
the period \(\Tso\) \citeApp{app:td-smes}.

\section{Conclusions}
\label{sec:conclusions}

In conclusion, we have demonstrated that we can generate
spin-polarized electrons, even when ionizing using linearly polarized
light, as long as we detect the photoelectrons in coincidence with the
ion. Furthermore, due to the recollision mechanism in strong-field
ionization, we are also able to control the spin of the photoelectron,
by tuning the ratio of the spin--orbit splitting and the angular
frequency of the driving field. This mechanism has important
implications for recollision-based imaging techniques such as
laser-induced electron diffraction, which use the energy-- and
angle-resolved distribution of the photoelectron to infer the state of
the ion; through the spin--orbit interaction, the spin of the
photoelectron would reveal additional information on the entangled
photoion.

\begin{acknowledgments}
  SCM would like to thank Edvin Olofsson for illuminating discussions.
  The work of SCM has been supported through scholarship 185-608 from
  \emph{Olle Engkvists Stiftelse}. JMD acknowledges support from the
  \emph{Knut and Alice Wallenberg Foundation} (2017.0104 and
  2019.0154), the Swedish Research Council (2018-03845) and \emph{Olle
    Engkvists Stiftelse} (194-0734). MI acknowledges support from
  \emph{Horizon 2020 research and innovation} (899794). OS
  acknowledges support from \emph{Horizon Europe} ERC-2021-ADG
  (\href{https://doi.org/10.3030/101054696}{101054696 Ulisses}).
\end{acknowledgments}

\bibliographystyle{apsrev4-2}
\bibliography{\bibliographyfile}

\appendix

\section{Methods}
\label{app:methods}

\subsection{Coulomb Repulsion}
\label{app:coulomb-repulsion}

The \emph{direct} and \emph{exchange interaction} potentials are
defined by their action on a spin-orbital
\begin{equation}
  \begin{aligned}
    \direct[cd]\ket{e} &\defd
                         \orbital{e}(\spatialspin_1)
                         \int\frac{\diff{\spatialspin_2}}{\abs{\vec{r}_1-\vec{r}_2}}
                         \conj{\orbital{c}}(\spatialspin_2)
                         \orbital{d}(\spatialspin_2), \\
    \exchange[cd]\ket{e} &\defd
                           \orbital{d}(\spatialspin_1)
                           \int\frac{\diff{\spatialspin_2}}{\abs{\vec{r}_1-\vec{r}_2}}
                           \conj{\orbital{c}}(\spatialspin_2)
                           \orbital{e}(\spatialspin_2)
                           \equiv \direct[ce]\ket{d},
  \end{aligned}
\end{equation}
where \(\spatialspin_{1,2}\) refer to both the spatial and spin
coordinates.

\subsection{Scaling the Spin--Orbit Interaction}
\label{app:scaling-so}

The explicit form of the RECP \eqref{eqn:recp} is
\begin{equation}
  \label{eqn:recp-detailed}
  \begin{aligned}
    \pseudopotential(\vec{r})
    &=
      \scalarpotential(r) +
      \mu\sopotential_{\ell j}(r),
    \\[0.75em]
    \scalarpotential(r)
    &\defd
      -\frac{Q}{r} +
      \operator{V}_\ell(r)
      \proj[\ell], \\
    \operator{V}_\ell(r)
    &\defd
      \frac{1}{2\ell+1}
      [\ell \operator{V}_{\ell,\abs{\ell-\frac{1}{2}}}(r) +
      (\ell+1) \operator{V}_{\ell,\ell+\frac{1}{2}}(r)], \\
    \proj[\ell]
    &\defd
      \proj[\ell,\abs{\ell-\frac{1}{2}}] +
      \proj[\ell,\ell+\frac{1}{2}],
    \\[0.75em]
    \sopotential_{\ell j}(r)
    &\defd
      \frac{\Delta\operator{V}_\ell(r)}{2\ell+2}
      [\ell\proj[\ell,\ell+1/2] -
      (\ell+1)\proj[\ell,\ell-1/2]],\\
    \Delta\operator{V}_\ell(r)
    &\defd
      \operator{V}_{\ell,\ell+1/2}(r) -
      \operator{V}_{\ell,\ell-1/2}(r),
    \\[0.75em]
    \operator{V}_{\ell j}
    &\defd
      B_{\ell j}^k\exp(-\beta_{\ell j}^kr^2),
  \end{aligned}
\end{equation}
where \(Q=26\) is the residual charge, \(\proj[\ell j]\) is a projector
on the spin--angular symmetry \(\ell j\), and \(B_{\ell j}^k\) and
\(\beta_{\ell j}^k\) are numeric coefficients found by fitting to
multiconfigurational Dirac--Fock all-electron calculations of the
excited spectrum \cite{Dolg2011,Dolg2016}.

\begin{figure}[htb]
  \centering
  \includegraphics{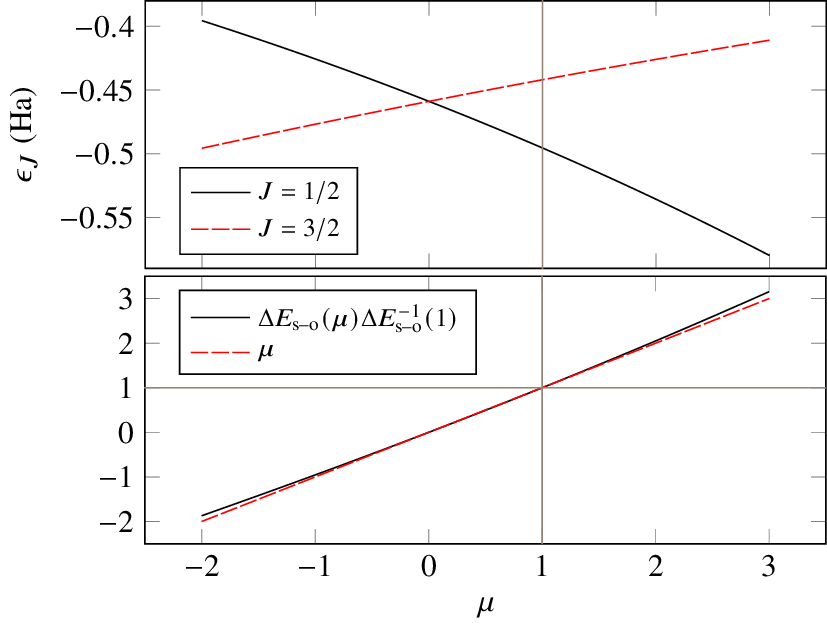}
  \caption{\label{fig:spin--orbit-splitting}Spin--orbit splitting
    \(\DEso(\mu) \defd \epsilon_{3/2}(\mu)-\epsilon_{1/2}(\mu)\)
    between \channela{} and \channelb{}, as a function of the scaling
    parameter \(\mu\) in Equation~\eqref{eqn:recp}.}
\end{figure}

In Figure~\ref{fig:spin--orbit-splitting}, we show the effect of
scaling the spin--orbit interaction in the RECP
\eqref{eqn:recp-detailed}. The resultant spin--orbit splitting is
essentially proportional to \(\mu\). In
Table~\ref{tab:ionization-energies}, the calculated ionization
potentials for the case \(\mu=1\) are compared with experimental values
from the literature.

\begin{table}[htb]
  \caption{\label{tab:ionization-energies} Calculated ionization
    potentials of the \(5\conf{s}\), \(\conf{5p}\) electrons of xenon,
    compared with their experimental and RCIS values (Refs:
    \cite{Saloman2004}\dataref{a}, \cite{Hansen1987PS}\dataref{b},
    \cite{Zapata2022}\dataref{c}).}
  \myruledtabular{l|S[table-format=2.3]S[table-format=2.4]S[table-format=2.3]S[table-format=2.4]}{%
    Hole & \mc{\unitlabel{\(\ionpotential\)}{\si{\electronvolt}}} &
    \mc{Exp.\ \unitbracket{\si{\electronvolt}}} &
    \mc{\unitlabel{\(\Delta\)}{\si{\electronvolt}}} &
    \mc{RCIS \unitlabel{\(\ionpotential\)}{\si{\electronvolt}}}
    \Bstrut\\
    \hline
    \Tstrut \channela & 12.026 & 12.130\dataref{a} & -0.104 & 11.968\dataref{c} \\
    \Tstrut
    \channelb & 13.483 & 13.436\dataref{b} & 0.047 & 13.404\dataref{c} \\
    \Tstrut
    \channelc & 27.927 & 23.397\dataref{b} & 4.530 & 27.485\dataref{c}}
\end{table}

\subsection{Photoelectron Spectra}
\label{app:photoelectron-spectra}
The photoelectron distributions, resolved on ion state \(I\),
photoelectron energy \(E\), the angle \(\theta\) with respect to the
polarization axis, and with spin projection \(\spin\), are obtained by
tracing over the reduced density matrix:
\begin{equation}
  \label{eqn:net-spin-spectrum}
  P_{I\spin}(E,\theta) \defd
  \int\diff{\phi}\matrixel{\spin}{\densitymatrix_{II}(E,\theta,\phi)}{\spin}.
\end{equation}

The density matrix is formed from the outer product of the
wavefunction with itself,
\begin{equation}
  \densitymatrix(E,\theta,\phi)
  \defd
  \ketbra{\Psi(\vec{k})}{\Psi(\vec{k})},
\end{equation}
using the close-coupling \emph{Ansatz} \cite{Fritsch1991} for the
wavefunction
\begin{equation}
  \label{eqn:wfn-close-coupling}
  \ket{\Psi(\vec{k})} =
  c_{I\vec{k}\spin}
  \antisym
  \ket{I}
  \ket{\vec{k}\spin}
\end{equation}
where \(\ket{I}\) is the state of the ion,
\begin{equation*}
  \vec{k}
  = \bmat{k\sin\theta\cos\phi\\
    k\sin\theta\sin\phi\\
    k\cos\theta},
  \quad
  k = \sqrt{2E},
\end{equation*}
is the asymptotic momentum of the photoelectron, \(\spin\) the spin
projection of the photoelectron, and \(\antisym\) the
antisymmetrization operator. The reduced density matrix is obtained
from the full density matrix, by projecting on specific ion states:
\begin{equation}
  \densitymatrix_{IJ}
  \defd
  \matrixel{I}{\densitymatrix}{J}.
\end{equation}
The decomposition \eqref{eqn:wfn-close-coupling} is computed
\cite{Carlstroem2022tdcisI, *Carlstroem2022tdcisII} using the tSURFF
\cite{Ermolaev1999, Ermolaev2000, Serov2001, Tao2012NJoP,
  Scrinzi2012NJoP} and iSURFV \cite{Morales2016-isurf} techniques.

\section{Classical Trajectories}
\label{app:classical-trajectories}

Here we re-derive the classical trajectories of a free electron in a
monochromatic electric field
\begin{equation*}
  \vec{F}(t) = \vec{F}_0\cos(\omega t);
\end{equation*}
these results have been presented many times, most notably by
\citet{Corkum1993PRL}.

We introduce the free oscillation range and the velocity amplitude:
\begin{equation*}
  \vec{\alpha}
  \defd
  \frac{\vec{F}_0}{\omega^2}; \qquad
  \vec{v}_0
  \defd
  \frac{\vec{F}_0}{\omega};
\end{equation*}
(we note that \(v_0^2=4\ponderomotive\)), as well as the phases
\(\phi=\omega t\), \(\phi_i=\omega t_i\), and \(\phi_r=\omega t_r\), \(\delta\phi=\phi_r-\phi_i\equiv\omega\tau\), etc.

We find the trajectories by integrating Newton's equations
\(\vec{a}(t) = -\vec{F}(t)\), neglecting the influence of the atomic
potential:
\begin{equation*}
  \vec{v}(t) =
  -\vec{F}_0
  \int_{t_i}^t\diff{t'}
  \cos\phi' =
  -\vec{v}_0
  (\sin\phi - \sin\phi_i),
\end{equation*}
\begin{equation*}
  \begin{aligned}
    \vec{r}(t)
    &=
      \vec{r}_0 +
      \vec{\alpha}
      [\cos\phi - \cos\phi_i +
      (\phi-\phi_i)
      \sin\phi_i].
  \end{aligned}
\end{equation*}
The phase of ionization \(\phi_i\) is found for each rescattering phase
\(\phi_r\) by requiring that the electron returns to the origin before
rescattering:
\begin{equation*}
  \vec{r}(t_r) = 0,
\end{equation*}
which we solve numerically using the gradient method.

The kinetic energy of the electron (before rescattering) is given by
\begin{equation}
  \label{eqn:kineng-classical}
  \kineng(t) =
  \frac{v^2(t)}{2} \equiv
  2\ponderomotive
  (\sin\phi - \sin\phi_i)^2.
\end{equation}

\begin{table}[htb]
  \caption{\label{tab:kinetic-energies-barrier}
    Influence of the initial position on the maximal kinetic energy
    upon rescattering, for \(F_0=\SI{3.5408e-2}{au}\) and
    \(\omega=\SI{0.0535}{\hartree}\).  }
  \myruledtabular{l|S[table-format=3.2]S[table-format=1.2]|S[table-format=1.2]S[table-format=1.2]S[table-format=1.2]}{%
    Barrier\Bstrut & {\unitlabel{\(r_0\)}{Bohr}} & {\unitlabel{\(\kinengmax\)}{\(\ponderomotive\)}} &
    {\unitlabel{\(\phi_i\)}{rad}} &
    {\unitlabel{\(\phi_r\)}{rad}} &
    {\unitlabel{\(\delta\phi\)}{rad}} \\
    \hline
    None\Tstrut & 0.0    & 3.17 & 0.31 & 4.40 & 4.08 \\
    Triangular & -12.96  & 4.69 & 0.68 & 4.27 & 3.59 \\
    Coulomb    & -10.49  & 4.36 & 0.59 & 4.30 & 3.71
  }
\end{table}
\begin{figure}[htb]
  \centering
  \includegraphics{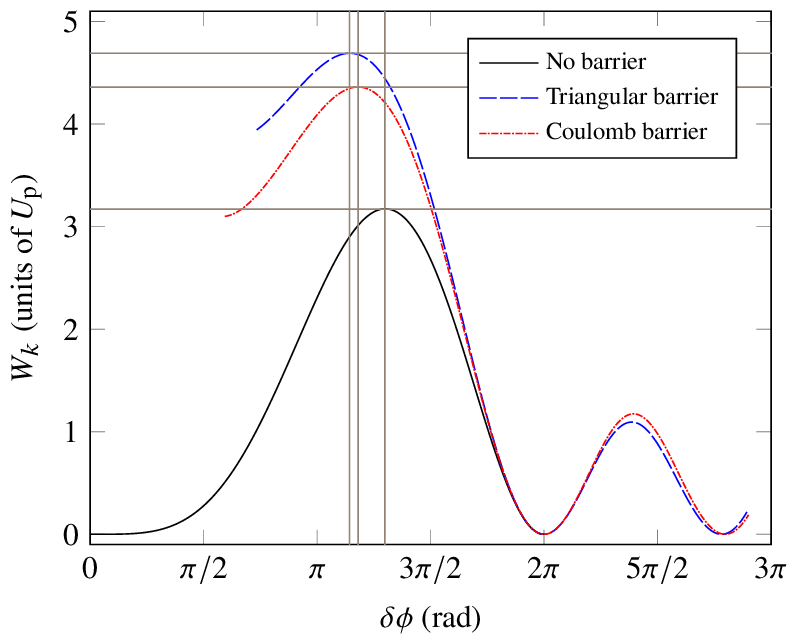}
  \caption{\label{fig:kinetic-energies-barrier} Kinetic energy upon
    rescattering as function of excursion phase, for three different
    choices of initial starting position (given in
    Table~\ref{tab:kinetic-energies-barrier}).}
\end{figure}
We may choose the initial position at the tunnel exit
\begin{equation}
  \begin{aligned}
    \rtriexit
    &= -\frac{\ionpotential}{F_0}, & \textrm{or} \\
    \rcoulexit
    &= -\frac{\ionpotential}{2F_0} -
      \sqrt{\left(\frac{\ionpotential}{2F_0}\right)^2 -
      \frac{2\ionpotential}{F_0}},
  \end{aligned}
\end{equation}
which will give maximal kinetic energies at the time of rescattering,
rather different from when the electron starts at the origin \(r_0=0\)
(see Table~\ref{tab:kinetic-energies-barrier} and
Figure~\ref{fig:kinetic-energies-barrier}), and in turn influence the
final momenta on the detector which upon rescattering had the maximal
kinetic energy (see Figure~\ref{fig:eyes-barrier}). Accounting for the
initial position is an important improvement compared to starting at
the origin as done by \citet{Spanner2004}, since it allows us to
correctly sample the off-axis spin-flip features as seen in
Figure~\ref{fig:spin-polarization-angle-resolved} of the main text;
for the figures shown there, we use the initial position
\(\rcoulexit\) for a Coulombic barrier.
\begin{figure}[htb]
  \centering
  \includegraphics{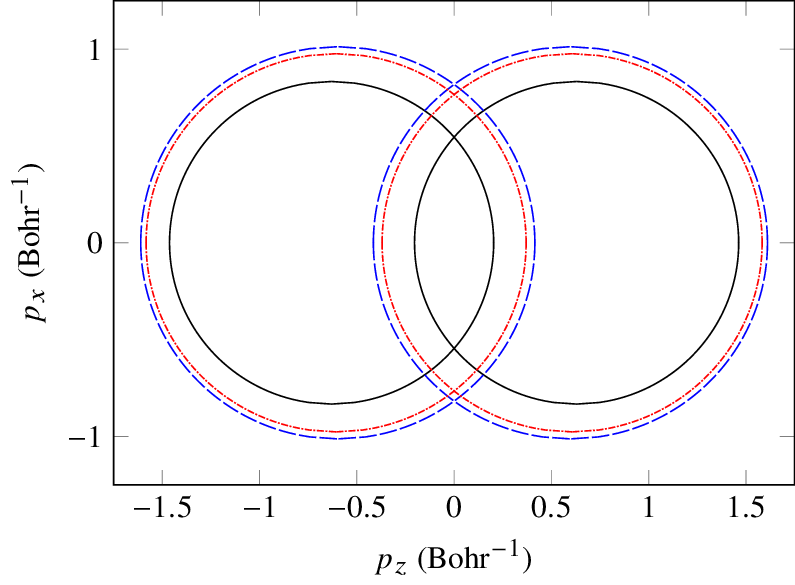}
  \caption{\label{fig:eyes-barrier} Final electron momenta
    corresponding to maximal kinetic energy upon rescattering, for
    three different choices of initial starting position (line
    patterns are the same as in
    Figure~\ref{fig:kinetic-energies-barrier}).}
\end{figure}

\subsection{Lineouts in the Backward Emission Direction}
\label{app:lineouts-backwards}

\begin{figure}[htb]
  \centering
  \includegraphics{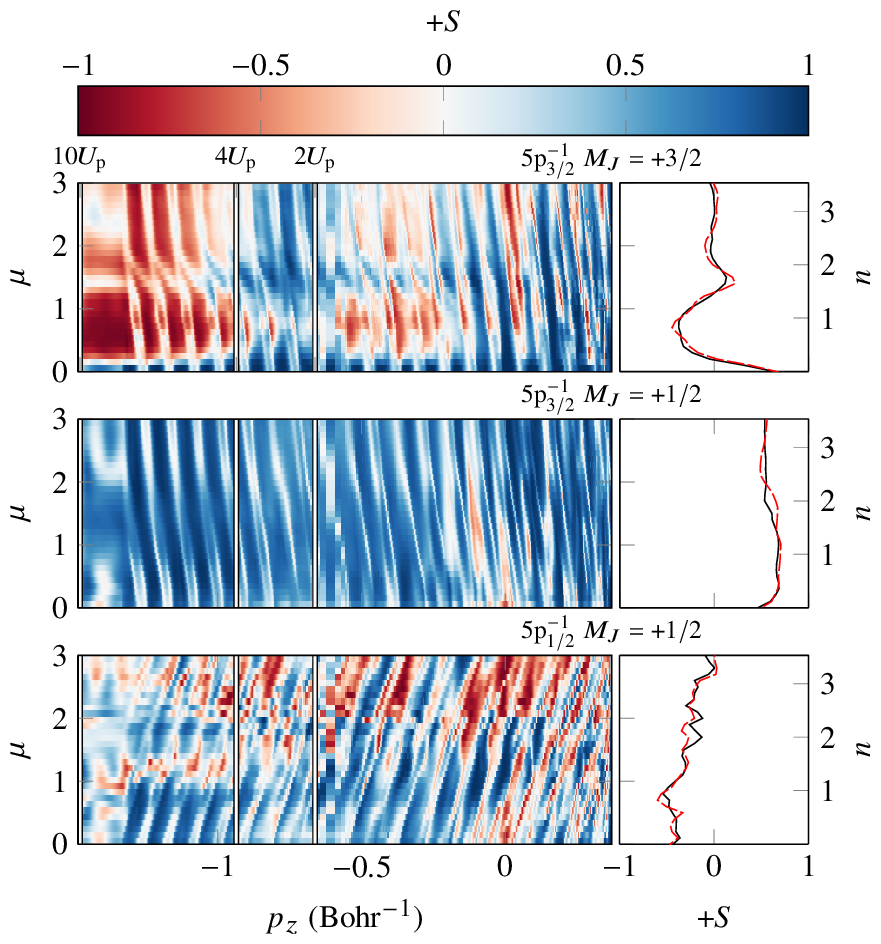}
  \caption{\label{fig:spin-polarization-3Up-lineouts-backward} Similar
    to Figure~\ref{fig:spin-polarization-3Up-lineouts} of the main
    manuscript, but the lineouts are instead along the \(\kinengmax\)
    circles in the backward direction, i.e.\ negative final \(p_z\).}
\end{figure}

In Figure~\ref{fig:spin-polarization-3Up-lineouts-backward}, the
lineouts along the \(\kinengmax\) circles in the backward emission
direction are shown. Due to the long pulse duration
(\(\tau=\SI{15}{\femto\second}\)), the spin polarization in the backward
direction is almost a perfect mirror image of the forward
distribution, as evidenced by the similarity of the integrated
lineouts also shown in the figure.

\section{Scattering Matrix Elements}
\label{app:smes}

\subsection{Effect of Removing \(\exchange[mm]\)}
\label{app:remove-Kmm}

\begin{figure}[htb]
  \centering
  \includegraphics{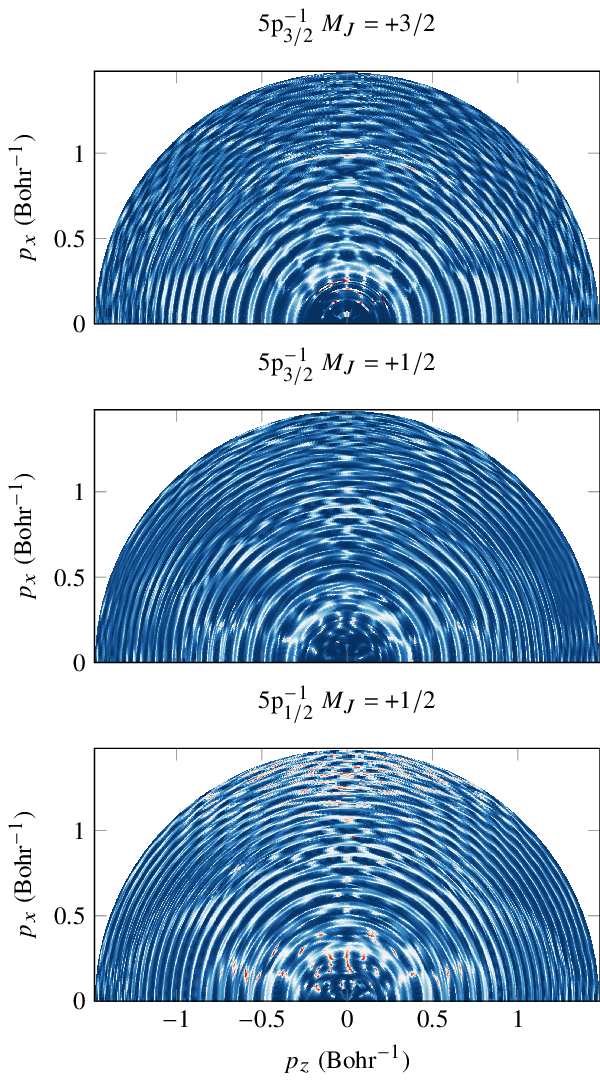}
  \caption{\label{fig:spin-polarization-kill-coulomb-Kmm} The effect
    of removing \(\exchange[mm]\) from the EOMs does not appreciably
    affect the spin-flipping mechanism;
    cf. Figure~\ref{fig:spin-polarization-kill-coulomb} of the main
    manuscript.}
\end{figure}

See Figure~\ref{fig:spin-polarization-kill-coulomb-Kmm} for the effect
of removing \(\exchange[mm]\) from the EOMs; the results do not change
appreciably.

\subsection{Time-Dependent Scattering Matrix Elements}
\label{app:td-smes}

Our numerical treatment is done in the \(jj\) coupling basis. We wish
to derive an expression for the time-dependent spin flip. The natural
basis for this process, the spin--orbit clock, is \(LS\) coupling,
where the spin of the hole is \enquote{breathing} in time, due to the
non-diagonal ionic Hamiltonian. For the \(n\conf{p^6}\)-multiplet, the
transform between \(LS\) and \(jj\) coupling is given by
\citep[see Table 8.1 of][]{Varshalovich1988}
\begin{equation}
  \begin{array}{rr|l|ll|ll|l}
    \Bstrut j & m_j & \conf{p_+\alpha} & \conf{p_+\beta} & \conf{p_0\alpha} & \conf{p_0\beta} & \conf{p_-}\alpha & \conf{p_-\beta} \\
    \hline
    \Tstrut\Bstrut \frac{3}{2} & \frac{3}{2} & 1 && && & \\
    \hline
    \Tstrut \frac{3}{2} & \frac{1}{2} & & \sqrt{\frac{1}{3}} & \sqrt{\frac{2}{3}}&&&\\
    \Bstrut \frac{1}{2} & \frac{1}{2} & & \sqrt{\frac{2}{3}} & -\sqrt{\frac{1}{3}}&&&\\
    \hline
    \Tstrut \frac{3}{2} & -\frac{1}{2} && && \sqrt{\frac{2}{3}} & \sqrt{\frac{1}{3}}&\\
    \Bstrut \frac{1}{2} & -\frac{1}{2} && && \sqrt{\frac{1}{3}} & -\sqrt{\frac{2}{3}}&\\
    \hline
    \Tstrut \frac{3}{2} & -\frac{3}{2} && && && 1
  \end{array}
\end{equation}
which clearly shows that the \(J=\frac{3}{2}\), \(M_J=\pm\frac{3}{2}\)
channels are spin-pure. Similarly, the ionic spin--orbit Hamiltonian
within this multiplet is in \(jj\) coupling
\begin{equation}
  \SpinOrbitHamiltonian^{(jj)} =
  \bmat{0\\
    &0\\
    &&-\DEso\\
    &&&0\\
    &&&&-\DEso\\
    &&&&&0},
\end{equation}
the propagator of which in \(LS\) coupling is given exactly by
\begin{equation}
  \exp[-\im \SpinOrbitHamiltonian^{(LS)} (t-t_i)] =
  \bmat{1\\
    &a & b \\
    &b & c \\
    &&&c & b \\
    &&&b & a \\
    &&&&&1},
\end{equation}
where
\begin{equation}
  \begin{aligned}
    a
    &\defd
      \frac{1}{3}(1+2\ce^{\im\phi}),
    &
      b
    &\defd
      \frac{\sqrt{2}}{3}(1-\ce^{\im\phi}),
    &
      c
    &\defd
      \frac{1}{3}(2+\ce^{\im\phi}),
  \end{aligned}
\end{equation}
and \(\phi=\DEso (t-t_i)\).

The matrix element responsible for the inelastic scattering between
channels is given by
\begin{equation}
  \label{eqn:sme}
  \twobodydx{l\cont{k}}{k\cont{l}} \equiv
  \twobody{l\cont{k}}{k\cont{l}} -
  \twobody{l\cont{k}}{\cont{l}k},
\end{equation}
the first term of which corresponds to the \emph{direct interaction},
and the second term to the \emph{exchange interaction}. As shown in
the above, when dropping \(\direct[lk]\) from the EOMs, the
spin-flipping mechanism was almost completely suppressed, which is why
we will focus on \(\twobody{l\cont{k}}{k\cont{l}}\), from which
\(\direct[lk]\) originates \cite{Carlstroem2022tdcisI}.

Assume we initially ionize the \(\ket{l(t_i)}=\ket{\conf{p_+\beta}}\)
orbital (a component of the \(j=3/2\), \(m_j=1/2\) orbital); then,
neglecting any effect of the spin--orbit interaction on the free
electron, \(\contket*{l}\) will be a \(\beta\) electron, while the
associated hole will evolve in time according to
\begin{equation}
  \ket{l(t)} =
  a
  \ket{\conf{p_+\beta}} +
  b
  \ket{\conf{p_0\alpha}}.
\end{equation}
Simultaneously, the channel we consider scattering \emph{into}, the
spin-pure \(j=3/2\), \(m_j=3/2\), has a time-independent hole, also in
\(LS\) coupling:
\begin{equation}
  \ket{k(t)} = \ket{\conf{p_+\alpha}}.
\end{equation}
From this, we deduce that the direct part of the scattering matrix
element \eqref{eqn:sme}, responsible for the apparent spin flip, is
\begin{equation}
  \label{eqn:tdsme}
  \begin{aligned}
    &
    \abs{\twobody{l\cont{k}}{k\cont{l}}}^2
    =
      \Big|
      a
      \underbrace{\twobody{\conf{p_+\beta};\cont{k}\spin}{\conf{p_+\alpha};\cont{l}\beta}}_{0} +
      b
      \twobody{\conf{p_0\alpha};\cont{k}\spin}{\conf{p_+\alpha};\cont{l}\beta}
      \Big|^2 \\
    &=
      \abs{\frac{\sqrt{2}}{3}(1-\ce^{\im\phi})}^2
      \abs{\twobody{\conf{p_0\alpha};\cont{k}\spin}{\conf{p_+\alpha};\cont{l}\beta}}^2
      \delta_{\spin\beta}\\
    &=
      \frac{4}{9}
      [1-\cos\DEso(t-t_i)]
      \abs{\twobody{\conf{p_0\alpha};\cont{k}\spin}{\conf{p_+\alpha};\cont{l}\beta}}^2
      \delta_{\spin\beta},
  \end{aligned}
\end{equation}
which will have its maximum when \(\DEso(t-t_i)=(2q+1)\pi\), i.e. odd
multiples of \(\pi\).

We would reach a similar conclusion, if we instead assumed ionization
to start from \(\ket{l(t_i)}=\ket{\conf{p_0}\beta}\).  This argument can
trivially be extended to the exchange interaction
\(\twobody{l\cont{k}}{\cont{l}k}\), and hence also
\(\abs{\twobodydx{l\cont{k}}{k\cont{l}}}^2\). As a side-note, since
the orbitals in the first coordinate of
\(\twobody{l\cont{k}}{k\cont{l}}\) in \eqref{eqn:tdsme} are both
\conf{p} electrons, only even orders in the multipole expansion of
\(\direct[lk]\) will contribute. Furthermore, since the orbitals have
different components (\(\conf{p_0}\) versus \(\conf{p_+}\)), the
lowest order is the quadrupole.

\end{document}